\documentclass[aps,10pt,prb,twocolumn,amssymb,amsmath,amsfonts,showpacs,
floatfix,citeautoscript,preprintnumbers,footinbib,superscriptaddress,a4paper
]{revtex4-1}
\usepackage{graphicx,subfigure,float,epstopdf}
\usepackage{amsmath,amsfonts,bm,graphicx,amssymb,color}
\usepackage{times,epstopdf}
\usepackage{graphicx}
\usepackage[latin1]{inputenc}
\graphicspath{{./Figures/}}
\usepackage[
   a4paper,
   dvips,
   breaklinks,
   ps2pdf,
   colorlinks=true,
   bookmarks=true,
   bookmarksopenlevel=3,
   bookmarksnumbered=true,
   menucolor=blue,
   linkcolor=blue,
   citecolor=blue,
   urlcolor=blue,
   pdfcreator={GhostScript},
   pdfkeywords={},
   pdfstartview=FitH]{hyperref}

\newcommand{\moy}[1]{\langle{#1}\rangle}

\begin{document}

\def\lptms{Laboratoire de Physique Th\'eorique et Mod\`eles statistiques, Universit\'e Paris-Sud, CNRS, UMR8626, 91405 Orsay, France.}

\title{Melting of a frustration-induced dimer crystal and incommensurability in the $J_1$-$J_2$ two-leg ladder}
\author{Arthur Lavar\'elo}\email{arthur.lavarelo@u-psud.fr}
\affiliation{\lptms}
\author{Guillaume Roux}\email{guillaume.roux@u-psud.fr}
\affiliation{\lptms}
\author{Nicolas Laflorencie}\email{nicolas.laflorencie@irsamc.ups-tlse.fr}
\affiliation{Laboratoire de Physique Th\'eorique, Universit\'e de Toulouse, UPS, (IRSAMC), Toulouse, France}

\date{\today}

\begin{abstract}
The phase diagram of an antiferromagnetic ladder with frustrating next-nearest neighbor couplings along the legs is determined using numerical methods (exact diagonalization and density-matrix renormalization group) supplemented by strong-coupling and mean-field analysis. Interestingly, this model displays remarkable features, bridging the physics of the $J_1$-$J_2$ chain and of the unfrustated ladder. The phase diagram as a function of the transverse coupling $J_{\perp}$ and the frustration $J_2$ exhibits an Ising transition between a columnar phase of dimers and the usual rung-singlet phase of two-leg ladders. The transition is driven by resonating valence bond fluctuations in the singlet sector while the triplet spin gap remains finite across the transition.
In addition, frustration brings incommensurability in the real-space spin correlation functions, the onset of which evolves smoothly from the $J_1$-$J_2$ chain value to zero in the large-$J_{\perp}$ limit. The onset of incommensurability in the spin structure-factor and in the dispersion relation is also analyzed. The physics of the frustrated rung-singlet phase is well understood using perturbative expansions and mean-field theories in the large-$J_{\perp}$ limit. Lastly, we discuss the effect of the non-trivial magnon dispersion relation on the thermodynamical properties of the system. The relation of this model and its physics to experimental observations on compounds which are currently investigated, such as BiCu$_2$PO$_6$, is eventually addressed.
\end{abstract}


\maketitle

Ladder materials offer a unique playground to improve our understanding of the subtleties arising from quantum effects in low dimensional geometries~\cite{Dagotto1996}. For antiferromagnetic $S=1/2$ Heisenberg models, it is well-known that ladders with an even number of legs $n_\ell$ display short-range correlations for any non-zero inter-chain coupling $J_{\perp}$ and a finite spin gap $\Delta_s\sim J_\perp\exp(-a n_\ell)$, whereas the gapless quasi-long-range ordered state of a single $S=1/2$ chain is robust for ladders having an odd number of legs~\cite{Affleck1988}. Ladder physics has been intensively explored during the last two decades both theoretically and experimentally, in particular regarding spin gap physics, impurity effects~\cite{Alloul2009}, field-induced magnetization processes~\cite{Giamarchi2008}, superconductivity in hole doped systems~\cite{Dagotto1996}, etc. Despite their rather simple geometry and the huge amount of studies, ladder systems remain a topic of current interest. Newly synthesized two-leg ladder materials which exhibit sizable spin gaps $\sim 1$ meV have recently emerged~\cite{Mentre06, Giamarchi2008}, thus opening the possibility to close the gap with an external magnetic field. 

\begin{figure}[b]
\centering
\includegraphics[width=0.8\columnwidth,clip]{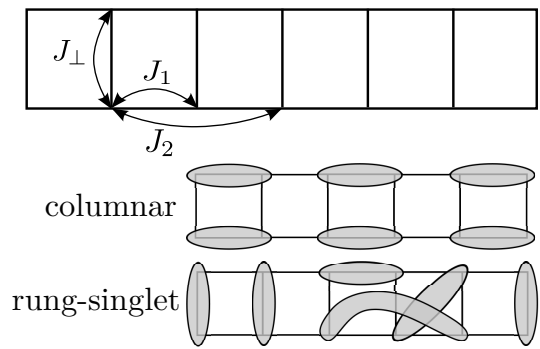}
\caption{The $J_1$-$J_2$ two-leg ladder model (top). Qualitative sketches of the two phases realized : the columnar dimer phase (middle) and the rung-singlet phase (down).}
\label{fig:model}
\end{figure}

The ground-state of a spin-$1/2$ Heisenberg two-leg ladder is a genuine quantum state which has no classical analog. Schematically, the short-range nature of the spin correlations can be encoded in the so-called resonating valence bond (RVB) picture with short range pairwise singlet bonds fluctuating over a few lattice sites~\cite{White1994}. Such a state, sometimes called ``rung-singlet'' (RS) because the strongest antiferromagnetic correlations are along the vertical rung bonds, is characterized by confined triplet excitations and a magnetic response to a local perturbation such as a non-magnetic impurity which remains confined in its vicinity, as probed by NMR. Another kind of spin-liquid which has also been intensively studied by many authors, back to Majumdar and Ghosh (MG) 40 years ago~\cite{Majumdar1969}, occurs the so-called $J_1-J_2$ chain. Used as a prototype for modeling the spin-Peierls transition in frustrated chains such as CuGeO$_3$, it displays a Kosterlitz-Thouless transition at a finite frustration ratio $J_2/J_1$ from a quasi-N\'eel-ordered state to a dimerized gapped state, which spontaneously breaks the (discrete) translational symmetry and is two-fold degenerate. The model under study in this manuscript bridges the physics of these two famous spin-liquids by considering two coupled $J_1$-$J_2$ chains, or equivalently, a two-leg ladder with frustrating couplings along the legs (see Fig.~\ref{fig:model}). It realizes a transition from a columnar dimer phase to the RS phase.

The search for a transition between the RS phase and dimerized phases in spin-$1/2$ two-leg ladders was initially triggered by Ref.~\onlinecite{Nersesyan1997} which proposed two scenarios. Based on the rewriting of the low-energy physics in terms of four gapped Majorana fermions fields~\cite{Shelton1996} comprising one in the singlet sector and three in the triplet sector. Taking into account a four spin interaction~\cite{Nersesyan1997} leads to closing the triplet gap or the singlet gap while the other sector remains gapped. The first scenario was realized in the well-studied two-leg ladder model with ring exchange~\cite{Brehmer1999}, displaying a transition to staggered dimer phase. The second scenario has been realized following the initial proposal by taking into account four spin interaction with a negative coupling~\cite{Fath2001}. Restricting models to two-body exchanges only, the main frustrated models which have been investigated to stabilize dimerized phases are two-leg ladders with cross-coupled, zig-zag and inequivalent exchanges on plaquettes~\cite{Weihong1998}.
Other models considering explicit dimerization can naturally yield staggered and columnar dimer orders~\cite{Martin-Delgado1996}.
In the context of the confinement-deconfinement transition~\cite{Nersesyan2003}, the $J_1$-$J_2$ ladder model for which there is an experimental realization~\cite{Volkova2010}, was studied using the low-energy physics~\cite{Starykh04} and numerics, yet limited to the MG point~\cite{Capriotti2004, Vekua2006} $J_2/J_1 = 0.5$ and with little evidence.

Here, we present an exhaustive study of the full phase diagram of the $J_1$-$J_2$ ladder model of Fig.~\ref{fig:model}, motivated by several aspects such as the shape of the whole phase diagram, the nature of the excitations, and the evolution of the incommensurability upon increasing frustration. Besides the mere theoretical interest, this model is also relevant to the spin ladder material BiCu$_2$PO$_6$ where several studies have pointed towards a scenario involving a sizable intra-chain frustration $J_2/J_1$~\cite{Koti07, Mentre09, Tsirlin10}. In particular, inelastic neutron scattering experiments display incommensurate spin correlations which originate from magnetic frustration~\cite{Mentre09}. Based on numerical and analytical calculations, we report a careful study of the incommensurate response upon increasing the frustrating coupling.

The rest of the paper is organized as follows. We first present in section~\ref{sec:model} the full phase diagram of the model which display two gapped phases (columnar dimer and rung-singlet) together with a crossover line which signals the onset of incommensurability. Numerical results are presented in section~\ref{sec:transition} where the nature of the quantum phase transition, low energy spectra and incommensurate correlations are studied. In section~\ref{sec:dispersion}, we present analytical results based on strong-coupling expansions at large $J_\perp$ and bond-operator mean-field approximation for the triplet excitation branch. They account for the evolution of the spin gap and incommensurate wave-vectors in the RS phase obtained by simulations. The temperature dependence of the magnetic susceptibility is presented in section~\ref{sec:thermo}. Finally, section~\ref{sec:conclusion} concludes the paper and discusses some experimental issues and open questions.

\section{Model and phase diagram}
\label{sec:model}

We study the following Hamiltonian on a spin-$1/2$ two-leg ladder and depicted on figure~\ref{fig:model}: $\mathcal{H} = \sum_{i=1}^L h_i$, with the three couplings 
\begin{align}
\label{eq:model-Jpara}
h_{i}=\quad  &J_{1}\left[\mathbf{S}_{i,1}\cdot\mathbf{S}_{i+1,1} + \mathbf{S}_{i,2}\cdot\mathbf{S}_{i+1,2}\right]\\
\label{eq:model-J2}
      + &J_{2} 	      \left[\mathbf{S}_{i,1}\cdot\mathbf{S}_{i+2,1} + \mathbf{S}_{i,2}\cdot\mathbf{S}_{i+2,2}\right]\\
\label{eq:model-Jperp}
      + &J_{\perp} 	  \,\mathbf{S}_{i,1}\cdot\mathbf{S}_{i,2} \;,
\end{align}
where $\mathbf{S}_{i,j}$ is the spin-$1/2$ operator acting at site $i$ of leg $j$ and the $J$s are the magnitude of the various couplings which are here taken to be antiferromagnetic ($J>0$). The zero-temperature phase diagram is derived using numerical methods well suited for frustrated systems : exact (or Lanczos) diagonalization (ED) and density-matrix renormalization group~\cite{White1992} (DMRG), keeping up to 2000 states in each block.
Calculations are carried out using periodic boundary conditions (PBC) for ED and open boundary conditions (OBC) for DMRG.
We now describe below the main features of the obtained phase diagram.

This model includes two limiting cases of two different spin-liquid phases which have been extensively studied. We summarize below the main known results about these case:
\begin{itemize}
\item[(i)] When $J_{\perp}=0$, the model corresponds to two decoupled $J_1$-$J_2$ chains. A single $J_1$-$J_2$ chain displays a transition between the gapless phase of the Heisenberg chain to a gapful dimerized phase for~\cite{Okamoto92} $J_{2,c}\simeq 0.24117\,J_1$. For the special Majumdar-Ghosh (MG) point~\cite{Majumdar1969} $J_2=0.5\,J_1$, the ground-state can be written in a simple product of dimers. The gap and the spin correlation length can be computed exactly at this point. Beyond this remarkable mathematical feature, the MG point physically turns out to be the point at which the correlation length is minimal and beyond which the real-space spin correlations start to be incommensurate. As incommensurability will be discussed in details in this paper, we already give the three definitions and notations of wave-vectors signaling incommensurability in the three different observables:
\begin{equation}
\begin{cases}
q &:\text{ real-space spin correlations}\\
\bar{q} &:\text{ maximum of the spin structure-factor}\\
q^* &: \text{ minimum of the magnon dispersion relation}
\end{cases}
\label{eq:qdef}
\end{equation}
In the $J_1$-$J_2$ chain, due to the finite correlation length, the spin structure-factor (Fourier transform of the real-space correlations) has a maximum at an incommensurate wave-vector $\bar{q}$ above the so-called Lifshitz point~\cite{Bursill1995} $J_{2,L}\simeq 0.5206\,J_1$. Notice that the onset of incommensurability in the minimum of the triplet dispersion relation at $q^*$ is expected to happen at a third different value of $J_2/J_1$ (as for the analogous spin-1 chain~\cite{Golinelli1999}), but remains hard to study numerically because of limited sizes with exact diagonalization. In the large $J_2/J_1$ limit of two weakly-coupled chains, the behaviors of the spin gap and $q$ follow reasonably well the bosonization predictions~\cite{White1996, Kumar2010}. 
\item[(ii)] When $J_{2}=0$, the model is that of two coupled Heisenberg chains, i.e. the  Heisenberg spin ladder. A spin gap in the magnetic excitation opens linearly with the transverse coupling~\cite{Barnes1993} $J_{\perp}$ with logarithmic corrections~\cite{Totsuka1995, Shelton1996} at small $J_{\perp}$, leading to the RS phase. The naming originates from the large-$J_{\perp}$ regime in which the picture of the ground-state boils down to dimers on each rung. However, the correct physical picture of the isotropic ladder is that of an RVB spin-liquid~\cite{White1994} rather than that of pinned dimers along rungs.
\end{itemize}

\begin{figure}[t]
\centering
\includegraphics[width=0.95\columnwidth,clip]{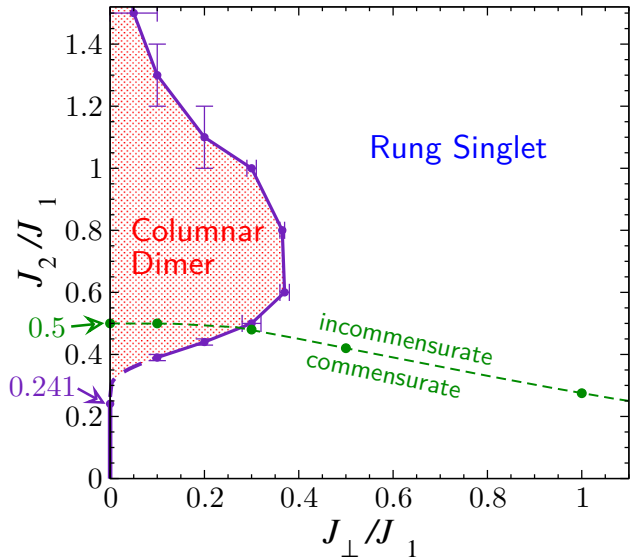}
\caption{(color online). Phase diagram of the  $J_1$-$J_2$ ladder. The commensurate-incommensurate regimes separated by the dashed green line corresponds to the incommensurability emerging in the real-space spin correlation functions. The violet transition line at finite $J_{\perp}$ is of the Ising type. The dashed violet line is the conjectured behavior of the transition line close to $J_{2,c} = 0.241\,J_1$.}
\label{fig:phasediag}
\end{figure}

The main result of the paper is the phase diagram of Fig.~\ref{fig:phasediag} which displays two main phases (apart from the critical line $J_{\perp}=0$ and $J_2<J_{2,c}$) : a columnar phase of dimer which breaks the translational symmetry and the RS phase. The columnar phase, first proposed by Vekua and Honecker~\cite{Vekua2006} along the MG line $J_2 = 0.5\,J_1$, emerges naturally as the combination of two dimerized chains with their dimer patterns in phase.
The physical picture of the transition is the following : starting from $J_{\perp}=0$, we have two dimerized chains between which the interchain antiferromagnetic coupling will favor RVB fluctuations leading to the alignment of their dimers (reminiscent of two-dimensional~\cite{Starykh04} columnar phases) and a lowering of the order. RVB fluctuations enhanced by larger $J_{\perp}$ eventually destroy the dimer order through an Ising second-order transition to the RS phase (as expected for a single-component order parameter, the transition corresponding to breaking a discrete $\mathbb{Z}_2$ symmetry). As we will see, the triplet gap remains finite across the phase diagram and on the transition line between these two distinct spin-liquid phases. This melting of the dimer crystal is driven by the low-energy fluctuations which lie in the singlet sector.

The transition line has a remarkable non-monotonic behavior showing a reentrance of the RS phase at small $J_{\perp}$ when increasing the frustration $J_2$. This shape can be understood on simple qualitative grounds : the order parameter and spin gap in the $J_1$-$J_2$ chain (along the $J_{\perp}=0$ line) typically have a steep increase above the transition point and then a slow decrease for large $J_2$. Considering $J_{\perp}$ as a perturbation, the larger the initial order, the larger $J_{\perp}$ is required to destroy it. This argument is all the more valid as the magnitude of $J_{\perp}$ along the transition does not exceed $0.4\,J_1$. 
When the order parameter is in the Ising scaling regime, we may use Eq.~\ref{eq:Ising} as a rough estimate of the transition line which yields : $J_{\perp,c} \propto [D_{\text{chain}}(J_2)]^8$, with $D_{\text{chain}}(J_2)$ the dimer order parameter of a single $J_1-J_2$ chain.
Furthermore, the numerics are particularly difficult when $J_{\perp}$ and the gaps are small, which corresponds to the behavior of the transition line close to $J_{2,c}\simeq 0.24 J_1$ and at large $J_{2}$. There, we may conjecture that the transition line qualitatively follows the behavior of the $J_{\perp}=0$ order parameter and should therefore exhibit an exponential-like opening $J_{\perp,c}\propto e^{-AJ_1/(J_2-J_{2,c})}$ close to $J_{2,c}$ and an exponential tail $J_{\perp,c}\propto e^{-A'J_2/J_1}$ at large $J_2/J_1$.

The last remarkable feature of this phase diagram is the onset of incommensurability which is a typical signature of frustration, already present at the classical level. As already recalled in the case of the frustrated spin chain, the onset of incommensurability depends on choosing the correlations in real or reciprocal spaces, or looking at the dispersion relation. In Fig.~\ref{fig:phasediag}, we chose the standard definition for the onset of incommensurability as the appearance of an incommensurate wave-vector in the real-space spin correlations. 

In the following sections, we explain how these results are obtained and describe in details the physics throughout the phase diagram using several observables.

\section{Nature of the transition}
\label{sec:transition}

\subsection{Order parameter}

\begin{figure}[t]
\centering
\includegraphics[width=\columnwidth,clip]{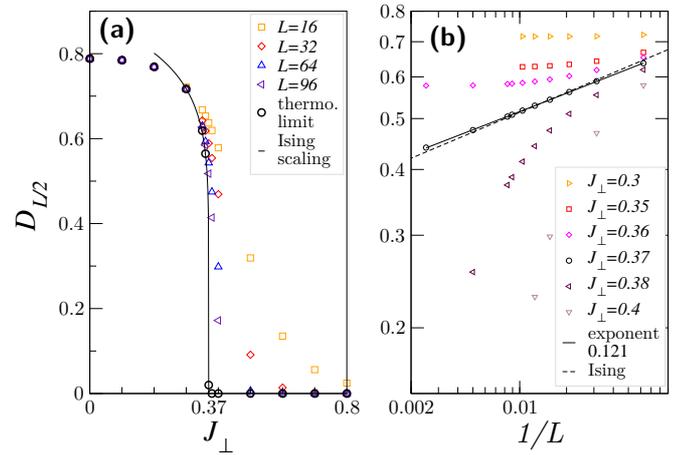}
\caption{(color online). \textbf{(a)} Behavior of the order parameter across the transition at fixed $J_2=0.6$. Extrapolations to the thermodynamic limit are performed using a polynomial fit in the ordered phase and Eq.~\eqref{eq:dimerfit} close to the transition point. \textbf{(b)} Scaling of the order parameter around and at the critical showing the exponent of the Ising universality class.}
\label{fig:order-transition}
\end{figure}

In this section, we give numerical evidence for the Ising nature of the transition between the columnar and RS phases. The simplest way to distinguish the two phases is to compute the dimer order parameter defined around site $i$ on leg $j$ by
\begin{equation}
D_{i,j} = \moy{\mathbf{S}_{i,j}\cdot\mathbf{S}_{i+1,j}}-\moy{\mathbf{S}_{i-1,j}\cdot\mathbf{S}_{i,j}}\;.
\label{eq:dimer-order}
\end{equation}
Using DMRG with OBC, one can directly access $D_{i,j}$ as a local order parameter. 
We check that the dimers are aligned on each legs and not staggered, which would be the other possible pattern on a ladder but energetically less favorable in this model. Therefore, we drop the leg index $j$ in the following.
Due to OBC, Friedel oscillations develop in the dimer order from the edges which decay exponentially in both the columnar and RS phases (away from the transition) and contain incommensurate oscillations in the incommensurate regime. Ignoring these oscillations to keep only the envelope of the Friedel oscillations, their typical behavior near the edges is expected to be~\cite{Lecheminant2002}
\begin{equation}
D_{x} \simeq D_{\infty} + A\frac{e^{-x/\xi_{\text{dimer}}}}{x^{\alpha}}
\label{eq:dimerfit}
\end{equation}
with $x$ the distance from the edge, $D_{\infty}$ the value of the order parameter in the thermodynamic limit (in the bulk of the ladder), $A$ a constant and $\xi_{\text{dimer}}$ the correlation length associated with the dimer fluctuations. The exponent $\alpha$ accounts for power-law corrections which are particularly relevant close to the transition. We expect $D_{\infty}$ to be zero in the RS phase and finite in the columnar dimer phase. We keep the discussion of $\xi_{\text{dimer}}$ for Sec.~\ref{sec:spingap} but we can already point out that it is a different length-scale from the usual spin correlation length $\xi_{\text{spin}}$ obtained from spin correlations. 

The typical evolution of the order parameter $D_{L/2}$ at the middle of the ladder with increasing $J_{\perp}$ and fixed $J_2=0.6\,J_1$ is shown on Fig.~\ref{fig:order-transition}(a). The vanishing of the order parameter at the transition is in qualitative agreement with the Ising prediction 
\begin{equation}
D_{\infty}(J_{\perp}) \propto (J_{\perp,c}-J_{\perp})^{1/8}\;,
\label{eq:Ising}
\end{equation}
in the vicinity of the critical point. In order to have a much more accurate determination of the critical point, as well as a test of the universality class, we use the finite-size scaling of the order parameter $D_{L/2}$ given by Eq.~\eqref{eq:dimerfit}. In the quantum Ising universality class, the correlations of the order parameter have a decay exponent $1/4$ at the critical point (for which $\xi_{\text{dimer}} = \infty$), which gives an exponent $\alpha = 1/8$ for the associated Friedel oscillations. Fig.~\ref{fig:order-transition}(b) shows the scaling behavior across the transition and the good agreement between the exponent found numerically and the Ising one.

\subsection{Entanglement entropy}

\begin{figure}[t]
\centering
\includegraphics[width=0.95\columnwidth,clip]{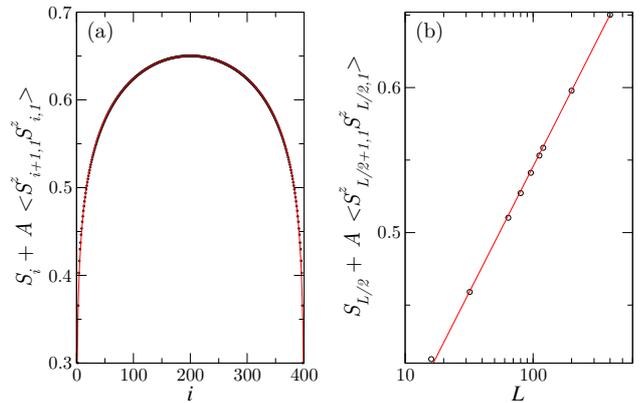}
\caption{(color online). (a) Entanglement entropy vs. block size $i$ fitted by Eq~\eqref{eq:EE-scaling} for the critical point of Fig.~\ref{fig:order-transition}. (b) Scaling of the entanglement entropy of a half-system with size $L$: The extracted central charge $c\simeq 0.46$ is in good agreement with the Ising value $c=1/2$.}
\label{fig:EE}
\end{figure}

Another supporting argument for the Ising nature of the transition is the value of the central charge $c=1/2$ at the critical point. In order to numerically extract $c$ from the DMRG data, we use the scaling of the entanglement entropy $S(x)$ with the half-block size $x$ for OBC:
\begin{equation}
S(x) = \frac{c}{6}\ln\left[\frac{L+1}{\pi}\sin\left(\frac{\pi x}{L+1}\right)\right] + A\moy{S^z_{x+1,1}S^z_{x,1}}+B,
\label{eq:EE-scaling}
\end{equation}
where $A$, $B$ and $c$ are fitting parameters. This formula is based on a universal prediction for the entanglement entropy~\cite{Calabrese2004} and subleading corrections~\cite{Laflorencie2006} $\moy{S^z_{x+1,1}S^z_{x,1}}$ which are here nothing but the local dimer order calculated numerically.
In order to obtain a good scaling, one must precisely locate the transition point (using the previous approach) and use rather large system sizes (here up to $L=400$). On Fig.~\ref{fig:EE}, we observe a scaling compatible with $c=1/2$, providing another independent evidence of the Ising nature of the transition. We stress that, since the spin correlations remain short-range at the critical point, the logarithmic increase of the entanglement entropy is only driven by the critical dimer fluctuations. As the entanglement entropy oscillations are related to the dimer order, its block-size dependence shows an incommensurate behavior as in Ref.~\onlinecite{Legeza2007}.

\subsection{Excitation spectra from ED}

\begin{figure*}[t]
\centering
\includegraphics[width=\textwidth,clip]{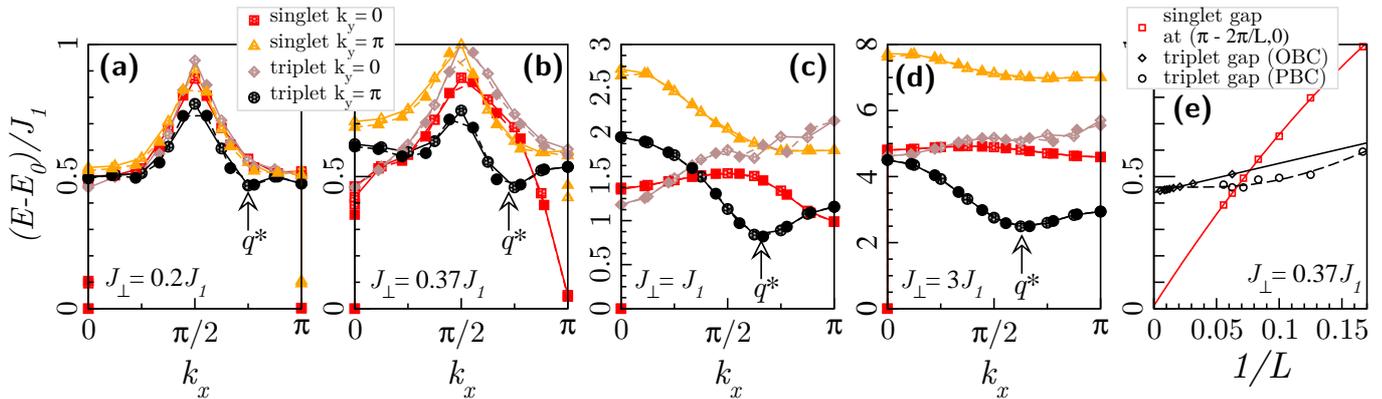}
\caption{(color online). \textbf{(a-d)} Low-energy excitations spectra for fixed $J_2=0.6\,J_1$ and increasing $J_{\perp}$. The up arrows indicate an approximate location of the incommensurate wave-vector $q^*$ corresponding to the minimum of the magnon relation dispersion. Full (dashed) symbols are for $L=18$ ($L=16$). \textbf{(e)} scaling of relevant gaps very close to the transition point : the triplet excitations remain gapped while a gapless singlet mode develops around the wave-vector $(\pi,0)$ as can be seen from the vanishing of the singlet gap at $(\pi-2\pi/L,0)$.}
\label{fig:excitation-spectra}
\end{figure*}

In this section, we discuss the nature of the low-energy excitations as a signature of the two phases. As a preamble, we recall that the deconfined spinons living on a dimerized chains must get confined as soon as the transverse coupling $J_{\perp}$ is turned on~\cite{Vekua2006}, since they must bind into a singlet state which size is controlled by $J_{\perp}$. Hence, the spinons excitations are confined throughout the phase diagram, except along the $J_{\perp}=0$ and $J_2>J_{2,c}$ line. 
We report in Fig.~\ref{fig:excitation-spectra} the finite-size spectra as functions of the momentum $k_x$ in the two parity sectors (defined by the transverse momentum $k_y$) and for the singlet and triplet channels (identified using spin inversion symmetry), for a fixed frustration $J_2=0.6\,J_1$ and increasing $J_{\perp}$. 

Starting with a small $J_{\perp}$ (Fig.~\ref{fig:excitation-spectra}(a)), the spectrum has four nearly degenerate low-energy states. They clearly stem from the four possible singlet states corresponding to the combination of two doubly degenerate chains. The degeneracy is lifted by $J_{\perp}$ which stabilizes the combinations associated with the columnar phase, while staggered configurations~\cite{Note1} have a gap controlled by $J_{\perp}$. Above these four low-energy singlet states, gapped singlet and triplet modes reminiscent of the dimerized chain~\cite{Shastry1981} develop around $k_x=0,\pi$. Their near degeneracy is a signature of weakling confined spinons. We notice a minimum in the magnon dispersion relation occurring at an incommensurate wave-vector $q^*\simeq 3\pi/4$, using the notation of Eq.~\eqref{eq:qdef}. Comparison between the two wave-vectors $q$ and $q^*$ will be addressed in Sec.~\ref{sec:incommensurability}.

Very close to the critical point (spectrum with $J_{\perp} = 0.37\,J_1$ of Fig.~\ref{fig:excitation-spectra}(b)), we observe that the triplet excitations remain fully gapped, while a low-energy singlet mode emerges around $(\pi,0)$ (the wave-vector of the nearly degenerate ground-state). The emergence of this mode is signaled in 
Fig.~\ref{fig:excitation-spectra}(e) by the scaling to zero of the singlet gap at $(\pi-2\pi/L,0)$. On the contrary, the triplet modes clearly have a finite gap in the thermodynamic limit. This singlet mode is related to the algebraic decay of the dimer correlations, supporting the melting of the dimer crystal scenario for the transition. This scenario seems to agree with the field theoretical approach and early numerical results of Ref.~\onlinecite{Vekua2006}.

Entering the RS phase by increasing further $J_{\perp}$, the ground-state is now non-degenerate and the spectrum fully gapped with an incommensurate magnon branch as the lowest excitation. For $J_{\perp}=J_1$ (Fig.~\ref{fig:excitation-spectra}(c)), this branch enters the continuum of many-magnons excitations for small and large $k_x$, but for large enough transverse coupling ($J_{\perp}=3J_1$ in Fig.~\ref{fig:excitation-spectra}(d)), the incommensurate magnon branch is well separated from the continuum (as usually found for Heisenberg ladders). This strong-coupling physics is well captured by the perturbative calculations of Sec.~\ref{sec:dispersion}. 

\subsection{Incommensurate wave-vector from spin correlations}
\label{subsec:incommensurate-q}

The onset of incommensurability across the phase diagram is obtained by choosing the definition from real-space spin correlations, i.e. when $q$ deviates from $\pi$.
Following the approach of White and Affleck for the $J_1$-$J_2$ chain~\cite{White1996a}, we extract the incommensurate wave-vector $q$ with the following ansatz 
for the exponential decay of the spin correlator : 
\begin{equation}
\moy{(S_{y+x,1}^z-S_{y+x,2}^z)(S_{y,1}^z-S_{y,2}^z)} \propto \cos(qx+\varphi)\, \frac{e^{-x/\xi_{\text{spin}}}}{\sqrt{x}}\;,
\label{eq:correlations}
\end{equation}
where $\xi_{\text{spin}}$ is the spin correlation length, which is also extracted from the fit. The antisymmetric combination of the spin operators is chosen to get the signal corresponding to the magnon mode with $k_y=\pi$.
The magnitude of the frustrating term required to drive $q$ away from $\pi$ is reported on Fig.~\ref{fig:phasediag} by the green dashed line. The line starts at the MG point $J_2/J_1 = 0.5$ for $J_{\perp}=0$ and asymptotically decreases towards $J_2/J_1=0$ in the large $J_{\perp}$ limit. There, an accurate description of the onset line, as well as the comparison between the wave-vectors \eqref{eq:qdef} is accessible (see Sec.~\ref{sec:incommensurability}). We lastly notice that for large $J_2/J_1$, $q$ naturally reaches $\pi/2$ since the model boils down to two weakly coupled ladders with a doubled unit cell.

\section{Magnon dispersion relation}
\label{sec:dispersion}

This section is dedicated to a more detailed study of the magnon excitation branch and its main features : effect of the frustrating term, spin gap, correlations lengths and the onset of incommensurability. The strong-coupling limit offers tractable analytical predictions that are compared with exact numerical results.

\subsection{Perturbative approaches at large $J_{\perp}$}

\subsubsection{Strong-coupling expansion}
\label{subsec:disprel}

A strong-coupling expansion can be carried out in the large-$J_{\perp}$ limit taking $J_1$ and $J_2$ as perturbations. These two terms lead to nearest neighbor and next-nearest neighbor hoppings for the magnons and the dispersion relation reads
\begin{equation}
\omega(k) = J_{\perp} + J_1\cos(k) + J_2\cos(2k)+\frac{3}{4}\frac{J_1^2+J_2^2}{J_{\perp}}\;.
\label{eq:strong-coupling}
\end{equation}
In comparison to the standard ladder dispersion relation, frustration brings the harmonic $2k$ leading to incommensurability. Remarkably, and contrary to the case of a single $J_1$-$J_2$ chain, we here have an exact estimate for the wave-vector minimizing the magnon gap, which we wrote $q^*$ and which is, in principle, different from $q$. In this limit, if $J_2\leq J_1/4$, the minimum of the dispersion remains at $q^*=\pi$ while for $J_2 \geq J_1/4$, the minimum takes place at the incommensurate wave-vector
\begin{equation}
q^* = \arccos\left(-\frac{J_1}{4J_2}\right)\;,
\label{eq:inc-vec}
\end{equation}
which reaches $\pi/2$ in the large-$J_2$ limit. Note that in the regime where $J_1\ll J_{\perp},J_2$, the picture is that of two unfrustrated two-leg ladders with a doubled lattice spacing and couplings $J_2$,$J_{\perp}$, slightly coupled in a zig-zag way by the perturbation $J_1$. The doubling of the unit cell naturally gives a minimum at $q^*=\pi/2$ in the original lattice model. The threshold and incommensurate wave-vector \eqref{eq:inc-vec} are the ones that actually appear as the angle of the helical structure of the classical version of both the $J_1$-$J_2$ chain and ladder models, hence in the spin-wave calculation. While for a single chain and the small-$J_{\perp}$ regime, quantum fluctuations strongly change the onset value and the $q^*(J_2)$ function, the calculation is here exact in the large-$J_{\perp}$ regime.

\subsubsection{Bond-operator mean-field theory}
\label{subsec:BOMF}

The shape of the dispersion relation can be refined using bond-operator mean-field (BOMF) theory following Ref.~\onlinecite{Gopalan1994}. The starting point is here again the large-$J_{\perp}$ limit where the local Hilbert space on each rung is spanned by introducing a singlet creation operator $s_i^{\dagger}$, and three triplet creation operators $t_{i,x}^{\dagger}$, $t_{i,y}^{\dagger}$ and $t_{i,z}^{\dagger}$. The local constraint $s_i^{\dagger}s_i + \sum_{\alpha} t_{i,\alpha}^{\dagger}t_{i,\alpha}=1$ is introduced using a Lagrange multiplier which identifies to a local chemical potential $\mu$ which we assume to be uniform from translation invariance. The Hamiltonian terms \eqref{eq:model-Jpara}-\eqref{eq:model-Jperp} can then be rewritten using these creation operators as:
\begin{align*}
h_i =& \quad \frac{J_{\perp}}{4}[-3s_{i}^{\dagger}s_{i} + t_{i,\alpha}^{\dagger}t_{i,\alpha}] \\
 & + \frac{J_1}{2}[t_{i,\alpha}^{\dagger}t_{i+1,\alpha}s_{i+1}^{\dagger}s_{i}+t_{i,\alpha}^{\dagger}t_{i+1,\alpha}^{\dagger}s_{i+1}s_{i}+\text{h.c.}] \\
 & + \frac{J_2}{2}[t_{i,\alpha}^{\dagger}t_{i+2,\alpha}s_{i+2}^{\dagger}s_{i}+t_{i,\alpha}^{\dagger}t_{i+2,\alpha}^{\dagger}s_{i+2}s_{i}+\text{h.c.}] \\
 & -\mu[s_{i}^{\dagger}s_{i}+t_{i,\alpha}^{\dagger}t_{i,\alpha}-1] + \ldots
\end{align*}
in which $i$ labels the rung, there is an implicit sum over $\alpha=x,y,z$, and the dots accounts for triplet-triplet interactions that we neglect according to Ref.~\onlinecite{Gopalan1994}. The resulting Hamiltonian is then studied under a mean-field approximation relying on the fact that the RS phase is dominated by singlets on rungs in the large-$J_{\perp}$ regime. We thus take $s_i\simeq \bar{s}$ which yields the Hamiltonian (in Fourier space) governing the dynamics of the triplets:
\begin{align*}
\mathcal{H}_{\text{MF}} =& L\Big(-\frac{3}{4}J_{\perp} \bar{s}^{2} - \mu\bar{s}^{2} + \mu\Big) \\
 & + \sum_{k,\alpha} \Lambda_{k}t_{k,\alpha}^{\dagger}t_{k,\alpha}+\Delta_{k}[t_{k,\alpha}^{\dagger}t_{-k,\alpha}^{\dagger}+t_{k,\alpha}t_{-k,\alpha}]
\label{eq:MF-ham}
\end{align*}
with 
\begin{align*}
\Lambda_{k} &= J_{\perp}\bar{s}^2\Big(\lambda_{1} \cos(k) + \lambda_{2}\cos(2k)\Big) +\frac{J_{\perp}}{4} -\mu\\
\Delta_{k} &= \frac{J_{\perp}}{2} \bar{s}^{2} \Big(\lambda_1\cos(k) + \lambda_{2}\cos(2k)\Big)\;,
\end{align*}
where we introduce two small parameters $\lambda_{1,2}=J_{1,2}/J_{\perp}$. This Hamiltonian is solved by a Bogoliubov transformation leading to 
\begin{align*}
\mathcal{H}_{\text{MF}} = E_0+\sum_{k,\alpha} \omega(k)\,\gamma_{k,\alpha}^{\dagger}\gamma_{k,\alpha}\;,
\end{align*}
where 
\begin{equation}
E_0 = -L\Big[\frac{3}{4}J_{\perp} \bar{s}^2 + \mu(\bar{s}^2-1)\Big] + \frac{1}{2}\sum_{k} \omega(k)\;,
\label{eq:E0-MF}
\end{equation}
and $\omega(k)^2=\Lambda_k^2 - 4\Delta_k^2$. The two mean-field parameters $\mu$ and $\bar{s}$ are found by minimizing Eq.~\eqref{eq:E0-MF} which gives
\begin{eqnarray*}
\mu &=&-\frac{3}{4}J_{\perp}+\frac{J_{\perp}}{2L} \sum_{k} \frac{\lambda_1 \cos(k)+\lambda_2\cos(2k)}{\sqrt{1+d_1\cos(k)+d_2\cos(2k)}}\;,\\
\bar{s}^2 &=& \frac{3}{2}-\frac{1}{4L} \sum_{k} \frac{2+d_1\cos(k) + d_2 \cos(2k)}{\sqrt{1+d_1\cos(k)+d_2\cos(2k)}}\;,
\end{eqnarray*}
with $d_1=\frac{2\lambda_1\bar{s}^2}{1/4-\mu/J_{\perp}}$ and $d_2=\frac{2\lambda_2\bar{s}^2}{1/4-\mu/J_{\perp}}$. The self-consistent relations are solved numerically to obtain the best $(\mu,\bar{s})$. The dispersion relation, within the mean-field approximation, reads
\begin{equation}
\omega(k) = \Big(\frac{J_{\perp}}{4}-\mu\Big)\sqrt{1 + d_1\cos(k) + d_2\cos(2k)}\;.
\label{eq:MF-dispersion}
\end{equation}
As $d_2/d_1 = J_2/J_1$, one obtains the same condition $J_2=J_1/4$ for the onset of incommensurability and the same incommensurate wave-vector given by \eqref{eq:inc-vec}. Taking the expansion of the mean-field equation in the $\lambda$s, one recovers the same first-order correction as in Eq.~\eqref{eq:strong-coupling} but the second order corrections are different :
\begin{equation*}
\begin{split}
\omega(k) = J_{\perp} & \big[1+\lambda_1\cos(k)+\lambda_2\cos(2k) + \frac{1}{4}(\lambda_1^2+\lambda_2^2)\\
&-\frac{1}{2}(\lambda_1\cos(k)+\lambda_2\cos(2k))^2\big]\;.
\end{split}
\label{eq:MF-expansion}
\end{equation*}
In particular, the spin gap to second order is not well reproduced.

\begin{figure}[t]
\centering
\includegraphics[width=.85\columnwidth,clip]{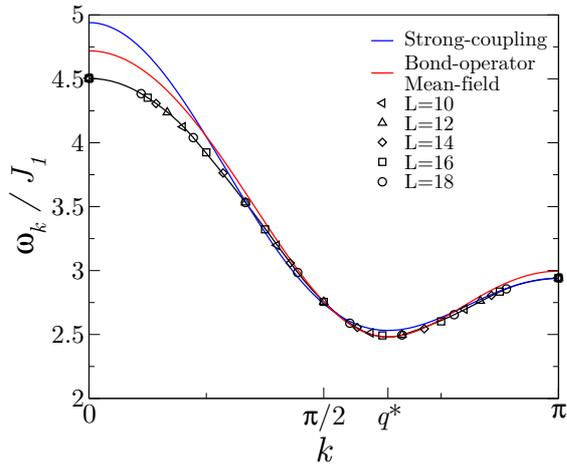}
\caption{(color online). Magnon dispersion relations for $J_\perp=3J_1$ and $J_2=0.6J_1$ from Lanczos diagonalization and compared to strong-coupling and BOMF theory.}
\label{fig:dispersion-relation}
\end{figure}

\subsubsection{Comparison with exact diagonalization}

We can now compare these analytical results to exact diagonalization results computed on ladders with up to $L=18$ rungs. As expected, Fig.~\ref{fig:dispersion-relation} shows a good quantitative agreement with the predictions in the strong-coupling regime in which finite-size effects are pretty small as one can judge from the data. The strong-coupling prediction gives an overall good account of the dispersion relation, with the best accuracy around $\pi$. The BOMF result \eqref{eq:MF-dispersion} needs a self-energy correction~\cite{Gopalan1994} of $0.7(J_1^2+J_2^2)/J_\perp^2$, to match the correct spin gap. Once this is done, the BOMF provides a better description than the strong-coupling expansion, particularly in the vicinity of $q^*$. Lastly, as for the well-known unfrustrated ladder, the picture from the strong-coupling limit remains qualitatively valid down to the isotropic ladder limit $J_1\simeq J_{\perp}$ (see Fig.~\ref{fig:excitation-spectra}).

\subsection{Spin gap and correlation lengths}
\label{sec:spingap}

\subsubsection{Spin gap and incommensurability with increasing frustration at large and fixed $J_{\perp}$}

\begin{figure}[t]
\centering
\includegraphics[width=\columnwidth,clip]{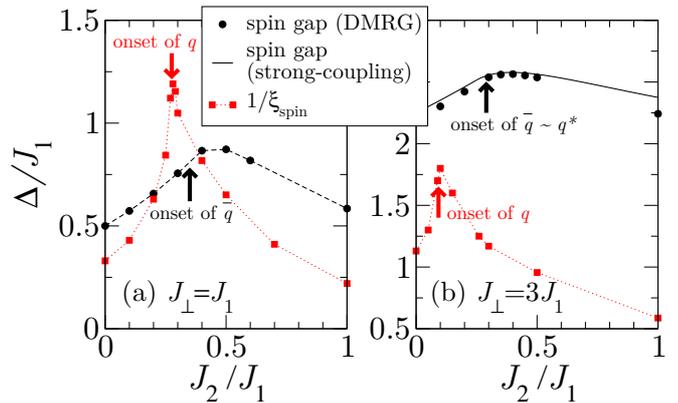}
\caption{(color online). Evolution of the triplet gap and inverse spin correlation length with $J_2$ at fixed $J_{\perp}$. The onset of the incommensurability in spin correlations ($q$), spin-structure factor ($\bar{q}$) and dispersion relation ($q^*$) are given when available. (a) the isotropic ladder situation $J_{\perp}=J_1$, where we have $\bar{q} \sim q$. The dashed line is just a guide for the eyes. (b) in the strong-coupling regime $J_{\perp}=3J_1$ where $\bar{q}$ and $q$ are very different. The line is the strong-coupling behavior of Eq.~\eqref{eq:spin-gap}.}
\label{fig:gap-vs-J2}
\end{figure}

A striking feature of the dispersion relation~\eqref{eq:strong-coupling} is that, at large-$J_{\perp}$, the gap first increases with $J_2$ before decreasing, hence passing through a local maximum. In the strong-coupling limit, the spin-gap $\Delta_s$ has the following behavior
\begin{equation}
\Delta_s = 
\begin{cases}
\displaystyle J_{\perp} - J_1 + J_2 + \frac{3}{4}\frac{J_1^2+J_2^2}{J_{\perp}} & \text{for }  J_2\leq J_1/4\\
\displaystyle J_{\perp} - J_2 - \frac{J_1^2}{8J_2} + \frac{3}{4}\frac{J_1^2+J_2^2}{J_{\perp}} & \text{for }  J_2\geq J_1/4
\end{cases}\;.
\label{eq:spin-gap}
\end{equation}
For $J_{\perp} = 3J_1$ in Fig.~\ref{fig:gap-vs-J2}(b), the agreement with the numerics is already pretty good. One may observe from Eq.~\eqref{eq:spin-gap} that the local maximum of the spin gap is actually realized for a value of $J_2$ larger than the onset of the $q^*$ incommensurability wave-vector. Neglecting terms in $J_2/J_{\perp}$, the maximum from \eqref{eq:spin-gap} occurs at $J_2 = J_1/\sqrt{8} \simeq 0.35\,J_1$, which agrees with Fig.~\ref{fig:gap-vs-J2}(b). For the isotropic ladder with $J_{\perp} = J_1$, the strong-coupling result is no more quantitative but the existence of a local maximum and its location above the onset of $q^*$ (not shown) are yet robust features (see Fig.~\ref{fig:gap-vs-J2}(a)). We will see in Sec.~\ref{sec:incommensurability} that $q^*$ gets very different from $q$ in the large-$J_{\perp}$ limit so that the local maximum is actually not related to the true onset of incommensurability associated with $q$.

One may wonder whether the spin correlation length $\xi_{\text{spin}}$ obtained from Eq.~\eqref{eq:correlations} has also a singular behavior as it has been observed for instance in the spin-1 chain~\cite{Schollwock1996, Nomura2003} and whether its location is related to the onset of incommensurability. From Fig.~\ref{fig:gap-vs-J2}, we do observe that $\xi_{\text{spin}}^{-1}$ displays a local singular maximum close to the onset of incommensurability associated with $q$ (and not $q^*$ or $\bar{q}$). Consequently, the spin correlation length is not minimal where the spin gap is maximal but before. These observations support that the physical onset of incommensurability is the one associated with $q$ since it is associated with a singular behavior in $\xi_{\text{spin}}$, as for the MG point of a single chain.

\subsubsection{Spin gap and correlation lengths for fixed $J_2$, increasing $J_{\perp}$}

\begin{figure}[t]
\centering
\includegraphics[width=0.8\columnwidth,clip]{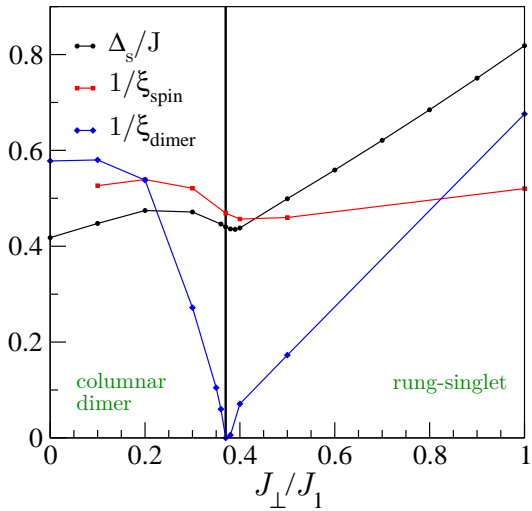}
\caption{(color online). Evolution of the spin gap and typical length scales with $J_{\perp}$ and a fixed $J_2=0.6\,J_1$. The vertical line materializes the critical point.}
\label{fig:gap-correlation length}
\end{figure}

We now turn to the behavior of the spin gap when crossing the transition line at a fixed $J_2=0.6\,J_1$ by increasing the transverse coupling $J_{\perp}$. 
The numerical results are displayed on Fig.~\ref{fig:gap-correlation length}. Although the spin gap always remains finite, it first decreases before increasing at large $J_{\perp}$. These behaviors are easily understood in two limiting cases : (i) at small $J_{\perp}$, the effect of the transverse coupling is essentially to destabilize the order parameter of the dimerized chains by RVB fluctuations and concomitantly slightly increases the spin gap, (ii) at large $J_{\perp}$, the spin gap is essentially controlled by the creation of a magnon on a rung, which increases with $J_{\perp}$ as found in the strong-coupling expression~\eqref{eq:spin-gap}. In between, there is actually a non-trivial local minimum found close to the transition but not exactly at it, and which is realized for a slightly larger value of $J_{\perp}$.

As regards the correlation lengths in the system, we exhibited previously two characteristic length scales $\xi_{\text{spin}}$ and $\xi_{\text{dimer}}$ associated with the spin and dimer fluctuations respectively. We do expect that these two length-scales behave very differently along this cut since we know that $\xi_{\text{dimer}}$ must diverge at the transition while $\xi_{\text{spin}}$, which is related to the inverse spin gap, must remain finite across the transition. This strong difference is illustrated on Fig.~\ref{fig:gap-correlation length} from which we observe nearly an order of magnitude between the two lengths around the critical point. Notice that $\xi_{\text{spin}}^{-1}$, as the spin gap, displays a minimum close to the transition but for a value of $J_{\perp}$ even larger than the one corresponding to the minimum of the spin gap.

\subsubsection{Evolution of the spin gap throughout the phase diagram}

\begin{figure}[t]
\centering
\includegraphics[width=0.8\columnwidth,clip]{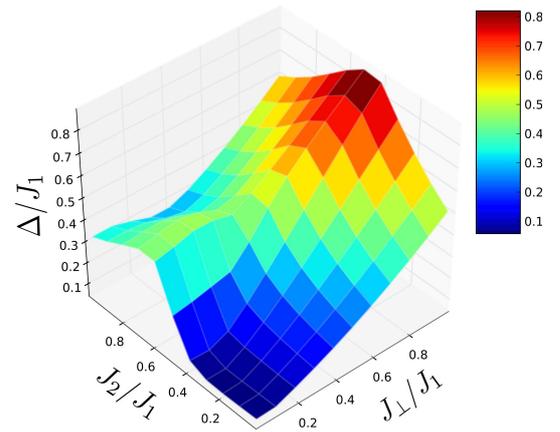}
\caption{(color online). Evolution of the triplet gap with $J_2$ and $J_{\perp}$ ($L=64$ sites).}
\label{fig:gap3D}
\end{figure}

From the two preceding cuts in the phase diagram, we observe that the interplay between transverse coupling and frustration results in non-monotonic behaviors of the spin gap throughout the phase diagram. How the local minimum observed close to the transition line interpolates with the local maximum arising in the vicinity of the onset of incommensurability in the dispersion relation is a challenging question. We give in Fig.~\ref{fig:gap3D} the typical evolution of the spin gap as a function of both $J_2$ and $J_{\perp}$ for a fixed size $L=64$.
The line of local maxima survives down to the weak-coupling regime where it meets the line of minima at a saddle point which is qualitatively close to the crossing point between the transition and incommensurability lines of the phase diagram of Fig.~\ref{fig:phasediag}. 
Another feature that is qualitatively well reproduced is the linear opening of the gap with $J_{\perp}$ (forgetting logarithmic corrections recalled in Sec.~\ref{sec:model}) starting from the critical phase below $J_{2,c}$. This region of very small gaps is demanding numerically and the exact behavior of the gap there is beyond the scope of this paper.
Finally, the tendency at large $J_2$ is a decrease of the gap for all $J_{\perp}$. Strictly speaking, in the limit $J_1=0$, as one recovers the situation of two decoupled ladders which couplings are $J_2$ and $J_{\perp}$, the gap is finite and starts to grow linearly with $J_{\perp}/J_2$. Only the corner point at $J_{\perp}=J_1=0$ and the critical line $J_2<J_{2,c},J_{\perp}=0$ have a zero gap.

\subsection{Onset of incommensurability}
\label{sec:incommensurability}

The onset and signatures of incommensurability is often a puzzling issue in frustrated models. As already introduced in Eq.~\eqref{eq:qdef}, we call $q$ the wave-vector of the real-space spin-correlators and $q^*$ the wave-vector at which the dispersion relation has its minimum. Having in mind that the spin structure factor $S(k)$ is accessible experimentally, we can further introduce the wave-vector $\bar{q}$ at which $S(k)$ is maximum. In unfrustrated antiferromagnets, the tendency to N\'eel ordering usually translates in the fact that $q=q^*=\bar{q}=\pi$. In the presence of frustration, we expect that, in general, the three wave-vectors can be different and that the value of the couplings at which they start to deviate from $\pi$ are not the same. For instance, in a gapped system, the difference between $q$ and $\bar{q}$ is easily understood from a Rayleigh criterion. Indeed, by taking the Fourier transform of short-range correlations at wave-vector $q$, one gets a double-peak structure for $S(k)$, which width is governed by $\xi_{\text{spin}}^{-1}$. We thus expect $\bar{q}$ to be larger than $q$, qualitatively $\bar{q} = q + \text{const.}\times \xi_{\text{spin}}^{-1}$. Clearly, the onset of $\bar{q}$ will occur after that of $q$ (when increasing frustration). Few studies have investigated the comparison between $q$ and $q^*$ among which we find Ref.~\onlinecite{Golinelli1999}.
 
\begin{figure}[t]
\centering
\includegraphics[width=0.9\columnwidth,clip]{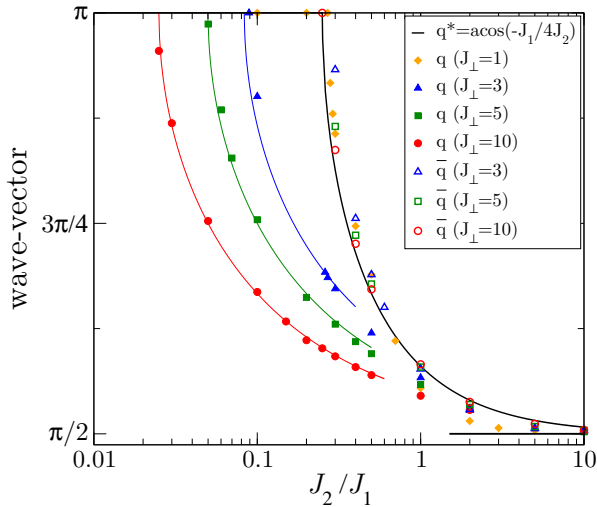}
\caption{(color online). Incommensurate wave-vectors $q$ and $\bar{q}$ with increasing frustration $J_2$ for different $J_{\perp}$ from DMRG results (dots) and compared with BOMF formula \eqref{eq:inc-vec-real} (straight lines). The black line gives the strong-coupling prediction for $q^*$ from Eq.~\eqref{eq:inc-vec}.}
\label{fig:incommensurate-q}
\end{figure}

Typical cuts at fixed $J_{\perp}$ and increasing $J_2$ are gathered in Fig.~\ref{fig:incommensurate-q}, showing that the larger $J_{\perp}$, the larger the discrepancy between the onset and the behavior of $q$ and $\bar{q}$. 
As discussed previously, all wave-vectors reach $\pi/2$ in the large-$J_2$ limit.
The black line recalls the prediction for $q^*$ from Eq.~\eqref{eq:inc-vec} for comparison. 
In particular, we observe that $\bar{q}$ is very close to $q^*$ in the strong-coupling regime. 

These observations can be explained by analytical calculations in the strong-coupling limit using a scenario proposed by Nomura~\cite{Nomura2003} for the onset of incommensurability. Since the one-magnon dispersion relation is in the $k_y=\pi$ sector, we only consider the structure factor in this sector which amplitude is much bigger than the one at $k_y=0$. We define the spin structure-factor as
\begin{equation*}
S(k)=S(k,\pi)=\sum_{x=1}^L e^{ikx}\moy{(S_{x,1}^z-S_{x,2}^z)(S_{1,1}^z-S_{1,2}^z)}\;.
\label{eq:structure-factor}
\end{equation*}
Within the BOMF approximation~\cite{Gopalan1994}, we obtain the following expression for $S(k)$ : 
\begin{equation}
S(k)=\bar{s}^2[1+d_1 \cos(k)+d_2 \cos(2k)]^{-\frac{1}{2}} \propto 1/\omega(k)
\label{eq:Sk-BOMF}
\end{equation}
where $d_1$ and $d_2$ are the two small positive parameters introduced in Sec.~\ref{subsec:BOMF}. Remarkably, since $S(k) \propto 1/\omega(k)$, incommensurability in $S(k)$ occurs on the same line $J_2=J_1 /4$ as for the dispersion relation, in other words $\bar{q}=q^*$. However, this is only true where the BOMF approximation is valid, that is, in the strong inter-chain coupling regime, and small deviations are found for smaller $J_{\perp}/J_1$ : for $J_{\perp}=J_1$, we have checked that the position of the structure factor maximum does not correspond to $q^*$ anymore (from Fig.~\ref{fig:excitation-spectra}).

\begin{figure}[t]
\centering
\includegraphics[width=\columnwidth,clip]{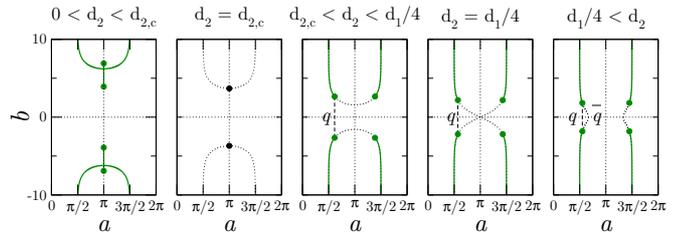}
\caption{(color online). Evolution of singularities (dots) and branch cuts (green lines) of the BOMF structure factor with increasing frustration. Dotted lines are the lines where $\Im[P(\cos k)]=0$.}
\label{fig:BranchCut}
\end{figure}

As shown on Fig.~\ref{fig:incommensurate-q}, incommensurability in $q$ occurs in the correlation function for smaller values of $J_2$ than in the structure factor. 
Indeed, oscillations in the real space should be connected to singularities of $S(k)$ in the complex plane~\cite{Schollwock1996, Nomura2003}, the maxima on the real axis being just a signature of the singularities. As described in Ref.~\onlinecite{Nomura2003}, the onset of incommensurability in the correlation function can be interpreted as a fusion of branch points in the complex plane. In order to discuss these singularities, the structure factor can be rewritten as
\begin{equation*}
S(k)=\dfrac{1}{\sqrt{P(\cos k)}}\ \ \text{with}\ \ P(X)=2d_2 X^2+d_1 X +1-d_2\,.
\end{equation*}
We set $k=a+ib$ with $a \in [0,2\pi]$ and $b \in \mathbb{R}$, which gives
\begin{align*}
P(\cos k)=&1+d_1\cos(a)\cosh(b)+d_2\cos(2a)\cosh(2b)\\
  &-i\sin(a)\sinh(b)\left[4d_2\cos(a)\cosh(b) + d_1\right]\,.
\end{align*}
On Fig.~\ref{fig:BranchCut}, we plotted the branch cuts and singularities of $S(k)$ in the complex plane $(a,b)$. The discriminant of $P$ changes sign for $d_2=d_{2,c}$ ;
\begin{equation}
d_{2,c}=\frac{1}{2}\Bigg(1-\sqrt{1-\frac{d_1^2}{2}}\Bigg)\simeq \frac{d_1^2}{8}\,.
\label{eq:d2c}
\end{equation}
For $d_2<d_{2,c}$, $P$ has two real roots both inferior to $-1$. As a result, $S(k)$ has four branch points on the axis $a=\pi$. For $d_2=d_{2,c}$, $P$ can be factorized and $S(k)$ has not any branch cuts but only two essential singularities on the axis $a=\pi$. For $d_2>d_{2,c}$, the roots of $P$ have a non-zero imaginary part, so the branch points of $S(k)$ are not on the axis $a=\pi$. Their real parts are given by the solutions of :
\begin{equation*}
2d_2 \cos^4(a)-(1+d_2)\cos^2(a)+\frac{d_1^2}{8d_2}=0\,,
\end{equation*}
that is $a=q$ or $2\pi-q$, with, after taking the strong-coupling approximation for $d_{1,2}$, the result :
\begin{equation}
q \simeq \arccos\left(-\frac{J_1}{2\sqrt{J_2J_\perp}}\right)\,.
\label{eq:inc-vec-real}
\end{equation}
This expression of the incommensurate wave-vector of the real-space correlations is in very good agreement with the DMRG calculations of Fig.~\ref{fig:incommensurate-q}. 
In particular, we see that the departure from $\pi$ for both $q$ and $q^*=\bar{q}$ has the same scaling exponent, with $q \sim \pi - A(J_2-J_{2,c})^{1/2}$, in agreement with the numerical findings of Ref.~\onlinecite{Schollwock1996}.
We also obtain the approximate value for the onset of incommensurability in $q$ as
\begin{equation}
J_2 = J_1^2 / 4J_{\perp}\;,
\label{eq:onset-q}
\end{equation}
showing that it vanishes at large $J_{\perp}$ in the phase diagram of Fig~\ref{fig:phasediag}, while the lines corresponding to $q^*$ and $\bar{q}$ would saturate at $J_2/J_1 = 1/4$.
To complete the discussion, we stress that the incommensurability emerges in $S(k)$ only when $d_2$ reaches $d_1/4$ (ie. $J_2=J_1/4$). This gives a graphical interpretation in the complex plane for the onset of incommensurability in $S(k)$ : the position of the maximum $\bar{q}$ is then given by the crossing of the line $\cos(a)\cosh(b) + d_1/(4d_2)=0$ (appearing in the imaginary part of $P$) with the real axis.

Looking back at Fig.~\ref{fig:incommensurate-q}, the situation for $J_{\perp}=J_1$ for which we find that $q$ is actually close to the strong-coupling prediction for $q^*$ is rather misleading and counter-intuitive. In fact, this qualitative agreement results from the fact that for this particular choice of $J_{\perp}$, the onset of $q$ is not very different from the approximate result of Eq.~\eqref{eq:onset-q} which itself numerically coincides with the onset of $q^*$. The difference between $q$ and $q^*$ is only seen for sufficiently large $J_2$, close to the asymptotic limit $\pi/2$. One may wonder whether $q$ gets locked to $\pi/2$ before the $J_1=0$ limit is attained, in a similar way as $q$ remains locked to $\pi$ below the incommensurability threshold. It is however hard to discriminate numerically between locking at a finite $J_2/J_1$ or an asymptotic approach to $\pi/2$ despite the seeming attraction by the $\pi/2$ line. Still, if there is a locking to $\pi/2$, we expect it to occur in the non-perturbative regime where $J_2\sim J_{\perp} \gg J_1$.

\section{Thermodynamics}
\label{sec:thermo}

In this section, we briefly address some issues related to the thermodynamics of the model. The magnetic susceptibility and specific heat vs. temperature can be computed from full diagonalization of small ladders, up to $L=10$. In the region of the phase diagram of Fig.~\ref{fig:phasediag} where energy scales are small and correlation lengths very large, we expect strong finite-size effects. We thus prefer to discuss the large-$J_{\perp}$ regime in which all gaps are sufficiently large. Still, before embarking to the strong-coupling regime, we can briefly comment on some interesting features of the thermodynamics of the columnar dimer phase. As we have seen from Fig.~\ref{fig:excitation-spectra}, there will be four low-lying singlet states which contributes to the specific heat but not to the susceptibility on finite clusters. More interestingly, right at the critical point, as a singlet gapless mode develops, we propose that the specific heat will have a linear behavior at low temperature (according to the Ising universal exponents), while the magnetic susceptibility will be exponentially suppressed with an energy scale given by the spin gap. The system size we have access to are two small to unambiguously demonstrate this appealing scenario. 

\begin{figure}[t]
\centering
\includegraphics[width=\columnwidth,clip]{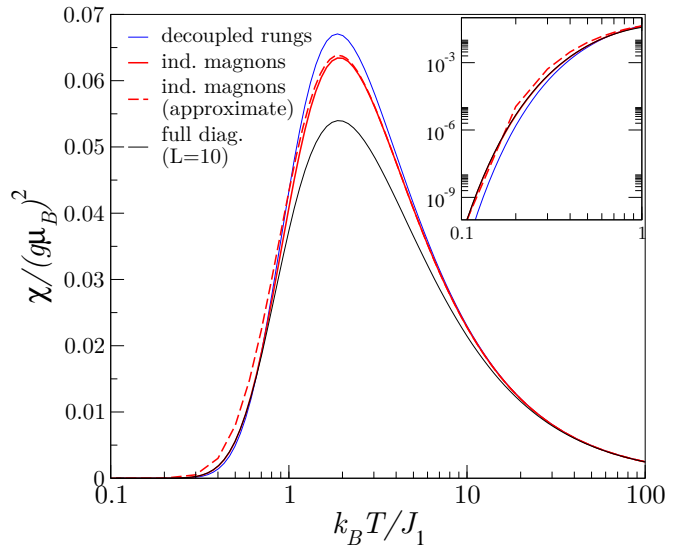}
\caption{(color online). Magnetic susceptibility $\chi$ (divided by $(g\mu_B)^2$ with $g$ the electronic dimensionless magnetic moment and $\mu_B$ the Bohr magneton) vs. temperature $T$ for $J_\perp=3J_1$ and $J_2=J_1/2$ from different methods. The inset displays the comparison at low temperatures on a log-log scale.}
\label{fig:susceptibility}
\end{figure}

We now discuss the behavior of the magnetic susceptibility in the rung-singlet phase. As we have reasonable analytical expressions for the magnon dispersion relation $\omega(k)$ in the strong interchain-coupling regime, we can derive the susceptibility assuming independent magnons, and compare it with numerical results. Using the approximate statistics of hard-core bosons proposed in Ref.~\onlinecite{Troyer1994}, one finds the following expression for the magnetic susceptibility :
\begin{equation}
\chi(\beta) = \beta \frac{z(\beta)}{1+3z(\beta)}\;,
\label{eq:susceptibility}
\end{equation}
with the one-magnon partition function $z(\beta)=\frac{1}{L}\sum_{k}e^{-\beta\omega(k)}$ where $\omega(k)$ is taken from Eq.~\eqref{eq:strong-coupling} and the inverse temperature $\beta=1/k_BT$. In Fig.~\ref{fig:susceptibility}, this ansatz is compared to the full diagonalization prediction and to the limit of decoupled rungs $J_1=J_2=0$ for a rather large $J_{\perp} = 3J_1$ and frustration $J_2 = 0.5J_1$. Since the gap from the strong-coupling dispersion relation is exact to second order in $J_{1,2}/J_{\perp}$, we predict the correct behavior at low temperature in which many-magnon effects, comprising both interaction and statistical effects, are negligible. On the opposite high-temperature limit, we recover the Curie law of decoupled rungs $\chi(\beta)\simeq\beta/4$ for all calculations. However, corrections from this Curie-law as observed in the full diagonalization results (finite-size effects are negligible at high temperature) are not quantitatively captured by the one-magnon dispersion relation, signaling the relevance of many-magnon effects mentioned above. At intermediate temperatures $T \approx J_1/k_B$ where the maximum of the susceptibility is found, the comparison is the worst, illustrating the fact that the value and the position of the maximum of the susceptibility is a non-trivial issue~\cite{Johnston2000}. 

One can however gain some qualitative information about the effect of frustration in the strong-coupling limit. 
The expression \eqref{eq:susceptibility} can be expanded in $J_{1,2}/J_{\perp}$ starting from the decoupled rung limit:
\begin{equation}
\begin{split}
\chi(\beta)\simeq&\; \chi_0(\beta) \times\\
&\left[ 1-\frac{1}{1+3e^{-\beta J_{\perp}}}\left(\dfrac{3}{4} \frac{J_{\text{eff}}}{J_{\perp}} (\beta J_{\text{eff}})
- \dfrac{1}{4} (\beta J_{\text{eff}})^2\right) \right]
\end{split}
\label{eq:susceptibility-SC}
\end{equation}
with $\chi_0(\beta) = \beta e^{-\beta J_{\perp}}/(1+3e^{-\beta J_{\perp}})$ the susceptibility of decoupled rungs, and the effective leg coupling $J_{\text{eff}} = \sqrt{J_1^2+J_2^2}$. The expansion reproduces well the independent magnon result down to temperatures below the maximum. When the temperature is too small ($\beta J_{\perp} \leq 1$), the expansion becomes questionable and the formula fails to reproduce the susceptibility. The interest of this formula is that despite its simplicity and strong approximations, it accounts for a non-trivial feature of the intermediate temperature regime which is the reduction of the maximum with $J_{\text{eff}}$. Since $J_{\text{eff}}$ is enhanced by frustration, we thus expect the maximum of the susceptibility to be sensitive to frustration.

\begin{figure}[t]
\centering
\includegraphics[width=\columnwidth,clip]{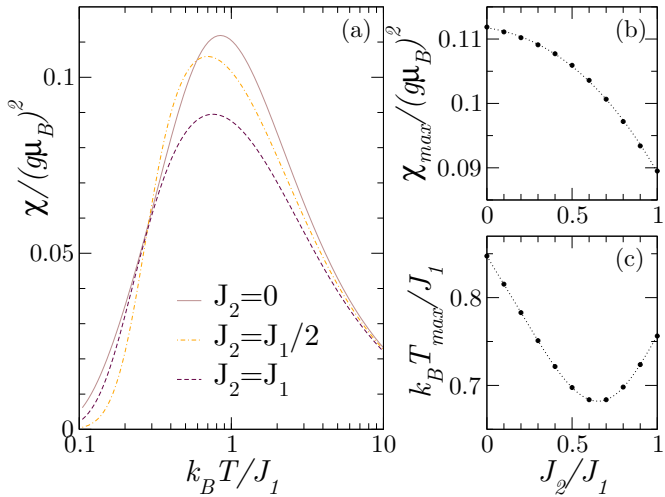}
\caption{(color online). (a) Evolution of the susceptibility with frustration $J_2$ for an isotropic ladder ($J_\perp=J_1$). (b) Maximum of the susceptibility with frustration. (c) Temperature of the maximum.}
\label{fig:susceptibilityVsJ2}
\end{figure}

In order to further investigate the effect of frustration in the RS phase, we plot in Fig.~\ref{fig:susceptibilityVsJ2}(a) the results of full diagonalization calculations for a fixed $J_{\perp}$ and increasing $J_2$. Two physical quantities govern the evolution of the curves : at low-temperature the spin gap sets an energy scale and we know from Fig.~\ref{fig:gap-vs-J2} that this gap has a non-monotonous behavior with $J_2$. Extracting the gap from low-temperature measurements can thus provide two possible values for $J_2$ and it is consequently not sufficient to use only the low-temperature regime. On the scale of Fig.~\ref{fig:susceptibilityVsJ2}(a), the change in the susceptibility due to the modification of the gap is visible. The intermediate and high-temperature regimes are rather controlled by $J_{\text{eff}}/J_{\perp}$ and show a systematic reduction of the susceptibility. As these two physical quantities evolves differently with$J_2$, the intermediate regime where the high and low temperature behaviors meet might have a non-trivial behavior. Indeed, the position of the maximum of the susceptibility displays a slightly non-monotonic evolution with $J_2$, qualitatively due to the evolution of the spin-gap (finite-size effects are small close to the maximum). High-temperature expansions~\cite{Oitmaa1996} can certainly capture the evolution of the maximum with $J_2$ but are beyond the scope of this article.

\section{Conclusion and discussion}

\subsection{Summary of main results}
\label{sec:conclusion}

In this paper, the phase diagram and most physically important properties of the $J_1$-$J_2$ two-leg ladder were computed. As often encountered in frustrated models, non-trivial physics is found : reentrance of the columnar dimer phase, low-energy RVB fluctuations, non-monotonic evolution of the spin gap, two characteristic lengths, and the onset of incommensurability. Apart from its experimental relevance which will be discussed in the next section, this model provides a simple physical picture and possibly experimentally relevant example for the melting of a dimer crystal. Furthermore, the surprising properties of the rung-singlet phase in the presence of frustration can be understood analytically and quantitatively in the large-$J_{\perp}$ regime. In particular, it provides a clear example and quantitative predictions on the difference between the different wave-vectors (denoted $q$, $\bar{q}$ and $q^*$ in the manuscript) signaling incommensurability. We eventually argue that the temperature-dependence of the magnetic susceptibility can be well understood in the low and large-$T$ limits, but that the intermediate regime where the maximum is found and its evolution with frustration cannot be quantitatively understood by approximate methods.

\subsection{Experimental relevance to BiCu$_2$PO$_6$}
\label{sec:experiments}

Frustration in 1D and quasi-1D quantum magnets plays a fundamental role. A famous example is the spin-Peierls transition observed almost two decades ago in the inorganic material CuGeO$_3$~\cite{Hase93}, believed to be a good realization of a $J_1-J_2$ chain with a frustrated ratio $J_2/J_1\sim 0.35$~\cite{Riera95}. More recently, a new class of frustrated ferromagnetic spin chains which display spiral (incommensurate) correlations and ferroelectric behavior, including LiCuVO$_4$, LiCu$_2$O$_2$ and Li$_2$ZrCuO$_4$ has been intensively studied~\cite{Masuda04}. 

Here, we want to focus on the spin ladder system BiCu$_2$PO$_6$~\cite{Mentre06,Koti07,Wang2010} which is believed to be a realization of the frustrated ladder model studied in this paper. So far, most of the studies agree on the fact that BiCu$_2$PO$_6$ has a ladder-like structure with a second-neighbor antiferromagnetic exchange along the leg direction~\cite{Koti07,Mentre09,Casola10,Tsirlin10}. However, various studies disagree regarding the amount of frustration $J_2/J_1$, the strength of the rung exchange $J_\perp$, and the inter-ladder couplings. Indeed, the size of the spin gap $\Delta_s/J_1\sim 0.3$ cannot be simply explained by an isolated frustrated ladder model where it is overestimated~\cite{Tsirlin10}. Additional 2D (and 3D) couplings~\cite{Bobroff09,Alexander10} between ladders are expected to play an important role: (i) to reduce the spin gap and (ii) to account for impurity-induced ordering at finite temperature~\cite{Bobroff09}. Another argument in favor of non-negligible inter-chain effects comes from the field-induced magnetization curve~\cite{Tsirlin10}. Taking all these facts into account, it is reasonable to argue that the most realistic model for the spin ladder material BiCu$_2$PO$_6$ should integrate both ingredients: in-leg frustration $J_2$ {\it{and}} higher-dimensional inter-ladder couplings. However, depending on the properties we want to investigate, dimensionality effect may matter or not. For instance, impurity-induced ordering at finite temperature is clearly a 3D effect~\cite{Bobroff09}, as well as NMR lines broadening at low temperature~\cite{Alexander10}. On the contrary, the incommensurate response is expected to be a 1D effect simply because the frustration is present only along the legs. Looking at Fig.~\ref{fig:incommensurate-q}, one can see that if $J_1\simeq J_\perp$ (which seems to be the case for BiCu$_2$PO$_6$), an incommensurate response is already expected for $J_2/J_1\sim 0.25$. 
Inelastic Neutron Scattering experiments on single crystal samples may allow for a direct evaluation of the incommensurate wave-vectors $\bar{q}$ and $q^*$ which should help to determine the frustrated character of this material.

The evaluation of the amount of frustration in BiCu$_2$PO$_6$ from susceptibility measurement requires to fit the full range of temperature to capture both the spin gap and the maximum $\chi_{\text{max}}$ of the susceptibility on equal footing. Inter-ladder interactions will also affect the spin gap and the many-magnons effects relevant in the intermediate and high temperature regimes. A simple unfrustrated model of coupled ladders has been shown using QMC simulations (unable to tackle frustrated models because of the sign problem) to correctly reproduce the low temperature part of the susceptibility while the maximum of $\chi$ was not correctly described~\cite{Alexander10}. Conversely, the high temperature part and the reduced value of $\chi_{\text{max}}$ was correctly captured by a 1D model including frustration (using ED) but at the expense of a spin gap largely overestimated~\cite{Tsirlin10}. Therefore, we believe that a precise experimental determination of the incommensurate response will be of crucial interest in order to estimate the amount of 1D frustration in this system. 

\acknowledgements

We thank Th. Jolicoeur, A. Kolezhuk and P. Lecheminant for illuminating discussions.

%\bibliographystyle{e-longmodern}
%\bibliography{LadderImpurity}

\begin{thebibliography}{10}

\bibitem{Dagotto1996}
E.~Dagotto and T.~M. Rice, Science {\bf 271}, 618 (1996).

\bibitem{Affleck1988}
I.~Affleck, Phys. Rev. B {\bf 37}, 5186 (1988).

\bibitem{Alloul2009}
H.~Alloul, J.~Bobroff, M.~Gabay, and P.~J. Hirschfeld, Rev. Mod. Phys. {\bf
  81}, 45 (2009).

\bibitem{Giamarchi2008}
T.~Giamarchi, C.~R{\"u}egg, and O.~Tchernyshyov, Nat. Phys. {\bf 4}, 198
  (2008).

\bibitem{Mentre06}
O.~Mentr\'e, E.~M. Ketatni, M.~Colmont, M.~Huv\'e, F.~Abraham, and V.~Petricek,
  J. Am. Chem. Soc. {\bf 128}, 10857 (2006).

\bibitem{White1994}
S.~R. White, R.~M. Noack, and D.~J. Scalapino, Phys. Rev. Lett. {\bf 73}, 886
  (1994).

\bibitem{Majumdar1969}
C.~K. Majumdar and D.~K. Ghosh, J. Math. Phys. {\bf 10}, 1388 (1969); J. Math. Phys. {\bf 10}, 1399 (1969).

\bibitem{Nersesyan1997}
A.~A. Nersesyan and A.~M. Tsvelik, Phys. Rev. Lett. {\bf 78}, 3939 (1997).

\bibitem{Shelton1996}
D.~G. Shelton, A.~A. Nersesyan, and A.~M. Tsvelik, Phys. Rev. B {\bf 53}, 8521
  (1996).

\bibitem{Brehmer1999}
S.~Brehmer, H.-J. Mikeska, M.~M{\"uller}, N.~Nagaosa, and S.~Uchida, Phys. Rev.
  B {\bf 60}, 329 (1999);
T.~S. Nunner, P.~Brune, T.~Kopp, M.~Windt, and M.~Gr\"uninger, Phys. Rev. B
  {\bf 66}, 180404 (2002);
M.~M\"uller, T.~Vekua, and H.-J. Mikeska, Phys. Rev. B {\bf 66}, 134423 (2002);
K.~Hijii and K.~Nomura, Phys. Rev. B {\bf 65}, 104413 (2002);
K.~Hijii, S.~Qin, and K.~Nomura, Phys. Rev. B {\bf 68}, 134403 (2003);
K.~P. Schmidt, H.~Monien, and G.~S. Uhrig, Phys. Rev. B {\bf 67}, 184413
  (2003);
A.~L\"auchli, G.~Schmid, and M.~Troyer, Phys. Rev. B {\bf 67}, 100409 (2003);
V.~Gritsev, B.~Normand, and D.~Baeriswyl, Phys. Rev. B {\bf 69}, 094431 (2004);
P.~Lecheminant and K.~Totsuka, Phys. Rev. B {\bf 71}, 020407(R) (2005).

\bibitem{Fath2001}
G.~F\'ath, O.~Legeza, and J.~S\'olyom, Phys. Rev. B {\bf 63}, 134403 (2001);
Y.-J. Wang, Phys. Rev. B {\bf 68}, 214428 (2003).

\bibitem{Weihong1998}
Z.~Weihong, V.~Kotov, and J.~Oitmaa, Phys. Rev. B {\bf 57}, 11439 (1998);
E.~H. Kim and J.~S\'olyom, Phys. Rev. B {\bf 60}, 15230 (1999);
E.~H. Kim, G.~F\'ath, J.~S\'olyom, and D.~J. Scalapino, Phys. Rev. B {\bf 62},
  14965 (2000);
H.-H. Hung, C.-D. Gong, Y.-C. Chen, and M.-F. Yang, Phys. Rev. B {\bf 73},
  224433 (2006);
B.~W. Ramakko and M.~Azzouz, Phys. Rev. B {\bf 76}, 064419 (2007);
E.~H. Kim, O.~Legeza, and J.~S\'olyom, Phys. Rev. B {\bf 77}, 205121 (2008);
G.-H. Liu, H.-L. Wang, and G.-S. Tian, Phys. Rev. B {\bf 77}, 214418 (2008).
T.~Hikihara and O.~A. Starykh, Phys. Rev. B {\bf 81}, 064432 (2010);
G.~Barcza, O.~Legeza, R.~M. Noack, and J.~Solyom, ArXiv e-prints, 1104.3990.

\bibitem{Vekua2006}
T.~Vekua and A.~Honecker, Phys. Rev. B {\bf 73}, 214427 (2006).

\bibitem{Martin-Delgado1996}
M.~A. Mart\'\i{}n-Delgado, R.~Shankar, and G.~Sierra, Phys. Rev. Lett. {\bf
  77}, 3443 (1996);
M.~A. Mart\'in-Delgado, J.~Dukelsky, and G.~Sierra, Physics Letters A {\bf
  250}, 430  (1998);
D.~C. Cabra and M.~D. Grynberg, Phys. Rev. Lett. {\bf 82}, 1768 (1999);
M.~J. Martins and B.~Nienhuis, Phys. Rev. Lett. {\bf 85}, 4956 (2000);
S.~J. Gibson, R.~Meyer, and G.~Y. Chitov, Phys. Rev. B {\bf 83}, 104423 (2011).

\bibitem{Nersesyan2003}
A.~A. Nersesyan and A.~M. Tsvelik, Phys. Rev. B {\bf 67}, 024422 (2003).

\bibitem{Volkova2010}
O.~Volkova, I.~Morozov, V.~Shutov, E.~Lapsheva, P.~Sindzingre, O.~C\'epas, M.~Yehia, V.~Kataev, R.~Klingeler, B.~B\"uchner, and A.~Vasiliev, Phys. Rev. B {\bf 82}, 054413 (2010).

\bibitem{Starykh04}
O.~A. Starykh and L.~Balents, Phys. Rev. Lett. {\bf 93}, 127202 (2004).

\bibitem{Capriotti2004}
L.~Capriotti, D.~J. Scalapino, and S.~R. White, Phys. Rev. Lett. {\bf 93},
  177004 (2004).

\bibitem{Koti07}
B.~Koteswararao, S.~Salunke, A.~V. Mahajan, I.~Dasgupta, and J.~Bobroff, Phys.
  Rev. B {\bf 76}, 052402 (2007).

\bibitem{Mentre09}
O.~Mentr\'e, E.~Janod, P.~Rabu, M.~Hennion, F.~Leclercq-Hugeux, J.~Kang,
  C.~Lee, M.-H. Whangbo, and S.~Petit, Phys. Rev. B {\bf 80}, 180413 (2009).

\bibitem{Tsirlin10}
A.~A. Tsirlin, I.~Rousochatzakis, D.~Kasinathan, O.~Janson, R.~Nath,
  F.~Weickert, C.~Geibel, A.~M. L\"auchli, and H.~Rosner, Phys. Rev. B {\bf
  82}, 144426 (2010).

\bibitem{White1992}
S.~R. White, Phys. Rev. Lett. {\bf 69}, 2863 (1992);
S.~R. White, Phys. Rev. B {\bf 48}, 10345 (1993);
U.~Schollw\"ock, Rev. Mod. Phys. {\bf 77}, 259 (2005).

\bibitem{Okamoto92}
K.~Okamoto and K.~Nomura, Physics Letters A {\bf 169}, 433  (1992);
G.~Castilla, S.~Chakravarty, and V.~J. Emery, Phys. Rev. Lett. {\bf 75}, 1823
  (1995);
S.~Eggert, Phys. Rev. B {\bf 54}, R9612 (1996).

\bibitem{Bursill1995}
R.~Bursill, G.~A. Gehring, D.~J.~J. Farnell, J.~B. Parkinson, T.~Xiang, and
  C.~Zeng, J. Phys.: Condens. Matter {\bf 7}, 8605 (1995).

\bibitem{Golinelli1999}
O.~Golinelli, T.~Jolicoeur, and E.~S{\"o}rensen, Eur. Phys. J. B {\bf 11}, 199
  (1999).

\bibitem{White1996}
S.~R. White, Phys. Rev. B {\bf 53}, 52 (1996).

\bibitem{Kumar2010}
M.~Kumar, Z.~G. Soos, D.~Sen, and S.~Ramasesha, Phys. Rev. B {\bf 81}, 104406
  (2010).

\bibitem{Barnes1993}
T.~Barnes, E.~Dagotto, J.~Riera, and E.~S. Swanson, Phys. Rev. B {\bf 47}, 3196
  (1993);
M.~Greven, R.~J. Birgeneau, and U.~J. Wiese, Phys. Rev. Lett. {\bf 77}, 1865
  (1996).

\bibitem{Totsuka1995}
K.~Totsuka and M.~Suzuki, J. Phys.: Condens. Matter {\bf 7}, 6079 (1995);
S.~Larochelle and M.~Greven, Phys. Rev. B {\bf 69}, 092408 (2004).

\bibitem{Lecheminant2002}
P.~Lecheminant and E.~Orignac, Phys. Rev. B {\bf 65}, 174406 (2002).

\bibitem{Calabrese2004}
P.~Calabrese and J.~Cardy, J. Stat. Mech. , P06002 (2004);
P.~Calabrese and J.~Cardy, J. Phys. A {\bf 42}, 504005 (2009).

\bibitem{Laflorencie2006}
N.~Laflorencie, E.~S. S\o{}rensen, M.-S. Chang, and I.~Affleck, Phys. Rev.
  Lett. {\bf 96}, 100603 (2006);
I.~Affleck, N.~Laflorencie, and E.~S. S\o{}rensen, J. Phys. A {\bf 42}, 504009
  (2009);
J.~Cardy and P.~Calabrese, J. Stat. Mech. {\bf 2010}, P04023 (2010).

\bibitem{Legeza2007}
O.~Legeza, J.~S\'olyom, L.~Tincani, and R.~M. Noack, Phys. Rev. Lett. {\bf 99},
  087203 (2007).

\bibitem{Note1}
The two possible staggered configurations have quantum numbers
  $(k_x,k_y)=(0,0)$ for the symmetric superposition (it lies in the same sector
  as the ground-state but just above), and $(k_x,k_y)=(\pi ,\pi )$ for the
  antisymmetric one.

\bibitem{Shastry1981}
B.~S. Shastry and B.~Sutherland, Phys. Rev. Lett. {\bf 47}, 964 (1981).

\bibitem{White1996a}
S.~R. White and I.~Affleck, Phys. Rev. B {\bf 54}, 9862 (1996).

\bibitem{Gopalan1994}
S.~Gopalan, T.~M. Rice, and M.~Sigrist, Phys. Rev. B {\bf 49}, 8901 (1994).

\bibitem{Schollwock1996}
U.~Schollw\"ock, T.~Jolicoeur, and T.~Garel, Phys. Rev. B {\bf 53}, 3304
  (1996).

\bibitem{Nomura2003}
K.~Nomura, J. Phys. Soc. Jpn. {\bf 72}, 476 (2003).

\bibitem{Troyer1994}
M.~Troyer, H.~Tsunetsugu, and D.~W\"urtz, Phys. Rev. B {\bf 50}, 13515 (1994).

\bibitem{Johnston2000}
D.~C. Johnston, R.~K. Kremer, M.~Troyer, X.~Wang, A.~Kl\"umper, S.~L. Bud'ko,
  A.~F. Panchula, and P.~C. Canfield, Phys. Rev. B {\bf 61}, 9558 (2000).

\bibitem{Oitmaa1996}
J.~Oitmaa, R.~R.~P. Singh, and W.~Zheng, Phys. Rev. B {\bf 54}, 1009 (1996).

\bibitem{Hase93}
M.~Hase, I.~Terasaki, and K.~Uchinokura, Phys. Rev. Lett. {\bf 70}, 3651
  (1993).

\bibitem{Riera95}
J.~Riera and A.~Dobry, Phys. Rev. B {\bf 51}, 16098 (1995).

\bibitem{Masuda04}
T.~Masuda, A.~Zheludev, A.~Bush, M.~Markina, and A.~Vasiliev, Phys. Rev. Lett.
  {\bf 92}, 177201 (2004);
H.~Katsura, N.~Nagaosa, and A.~V. Balatsky, Phys. Rev. Lett. {\bf 95}, 057205
  (2005);
M.~Enderle, C.~Mukherjee, B.~F\r{a}k, R.~K. Kremer, J.-M. Broto, H.~Rosner,
  S.-L. Drechsler, J.~Richter, J.~Malek, A.~Prokofiev, W.~Assmus, S.~Pujol,
  J.-L. Raggazzoni, H.~Rakoto, M.~Rheinst\"adter, and H.~M. R{\o}nnow,
  Europhys. Lett. {\bf 70}, 237 (2005);
M.~Mostovoy, Phys. Rev. Lett. {\bf 96}, 067601 (2006);
S.~Park, Y.~J. Choi, C.~L. Zhang, and S.-W. Cheong, Phys. Rev. Lett. {\bf 98},
  057601 (2007);
S.-L. Drechsler, O.~Volkova, A.~N. Vasiliev, N.~Tristan, J.~Richter,
  M.~Schmitt, H.~Rosner, J.~M\'alek, R.~Klingeler, A.~A. Zvyagin, and
  B.~B\"uchner, Phys. Rev. Lett. {\bf 98}, 077202 (2007);
N.~B\"uttgen, H.-A. Krug~von Nidda, L.~E. Svistov, L.~A. Prozorova,
  A.~Prokofiev, and W.~A\ss{}mus, Phys. Rev. B {\bf 76}, 014440 (2007);
F.~Schrettle, S.~Krohns, P.~Lunkenheimer, J.~Hemberger, N.~B\"uttgen, H.-A.
  Krug~von Nidda, A.~V. Prokofiev, and A.~Loidl, Phys. Rev. B {\bf 77}, 144101
  (2008);
S.~Seki, Y.~Yamasaki, M.~Soda, M.~Matsuura, K.~Hirota, and Y.~Tokura, Phys.
  Rev. Lett. {\bf 100}, 127201 (2008);
J.~Sirker, Phys. Rev. B {\bf 81}, 014419 (2010).

\bibitem{Wang2010}
S.~{Wang}, E.~{Pomjakushina}, T.~{Shiroka}, G.~{Deng}, N.~{Nikseresht}, C.~{R{\"u}egg}, H.~M.~{R{\o}nnow}, and K.~{Conder}, Journal of Crystal Growth {\bf 313}, 51 (2010).

\bibitem{Casola10}
F.~Casola, T.~Shiroka, S.~Wang, K.~Conder, E.~Pomjakushina, J.~Mesot, and H.-R.
  Ott, Phys. Rev. Lett. {\bf 105}, 067203 (2010).

\bibitem{Bobroff09}
J.~{Bobroff}, N.~{Laflorencie}, L.~K. {Alexander}, A.~V. {Mahajan},
  B.~{Koteswararao}, and P.~{Mendels}, Phys. Rev. Lett. {\bf 103}, 047201
  (2009).

\bibitem{Alexander10}
L.~K. Alexander, J.~Bobroff, A.~V. Mahajan, B.~Koteswararao, N.~Laflorencie,
  and F.~Alet, Phys. Rev. B {\bf 81}, 054438 (2010).

\end{thebibliography}

\end{document}